%Paper: hep-ph/9510420
%From: "E.W.N. Glover" <E.W.N.Glover@durham.ac.uk>
%Date: Thu, 26 Oct 95 15:26:23 GMT

\magnification=\magstephalf
\newbox\SlashedBox
\def\slashed#1{\setbox\SlashedBox=\hbox{#1}
\hbox to 0pt{\hbox to 1\wd\SlashedBox{\hfil/\hfil}\hss}{#1}}
\def\hboxtosizeof#1#2{\setbox\SlashedBox=\hbox{#1}
\hbox to 1\wd\SlashedBox{#2}}

% The following is necessary so that we can get a partial slash
% inside a math display... sigh.
\def\mathslashed#1{\setbox\SlashedBox=\hbox{$#1$}
\hbox to 0pt{\hbox to 1\wd\SlashedBox{\hfil/\hfil}\hss}#1}

\def\ifsmall{\iffalse}  % default is unreduced.
\def\titlepagefont{}  % default is ordinary font.

% the ps: landscape must be the first special command in order
% to get the first page in landscape mode -- so we go through some
% contortions to define TeXgraphics in the default case.
\def\DefineTeXgraphics{%
\special{ps::[global] /TeXgraphics { } def}}  % No need to do anything

\def\today{\ifcase\month\or January\or February\or March\or April\or May
\or June\or July\or August\or September\or October\or November\or
December\fi\space\number\day, \number\year}
\def\eatPrefix19{}
\def\Year{\expandafter\eatPrefix\the\year}
\newcount\hours \newcount\minutes
\def\monthname{\ifcase\month\or
January\or February\or March\or April\or May\or June\or July\or
August\or September\or October\or November\or December\fi}
\def\shortmonthname{\ifcase\month\or
Jan\or Feb\or Mar\or Apr\or May\or Jun\or Jul\or
Aug\or Sep\or Oct\or Nov\or Dec\fi}

\def\TimeStamp{\hours\the\time\divide\hours by60%
\minutes -\the\time\divide\minutes by60\multiply\minutes by60%
\advance\minutes by\the\time%
${\rm \shortmonthname}\cdot\if\day<10{}0\fi\the\day\cdot\the\year%
\qquad\the\hours:\if\minutes<10{}0\fi\the\minutes$}

%\DefineTeXgraphics}

%\DefineTeXgraphics}

%\DefineTeXgraphics}

\def\Title#1{%
\vskip 1in{\titlefont\centerline{#1}}\vskip .5in}
%\DefineTeXgraphics}

% restores pagenumbers

%\def\draft{\centerline{\it Preliminary Draft}\vskip 0.4in}

\newif\ifdraftmode
\newif\ifleftlabels  % Labels in left margins as well, for European-size paper

% Stolen from harvmac.tex 04/08/92
%       use \nolabels to get rid of eqn, ref, and fig labels in draft mode
\def\nolabels{\def\wrlabeL##1{}\def\eqlabeL##1{}\def\reflabeL##1{}}
\def\writelabels{\def\wrlabeL##1{\leavevmode\vadjust{\rlap{\smash%
{\line{{\escapechar=` \hfill\rlap{\sevenrm\hskip.03in\string##1}}}}}}}%
\def\eqlabeL##1{{\escapechar-1\rlap{\sevenrm\hskip.05in\string##1}}}%
\def\reflabeL##1{\noexpand\rlap{\noexpand\sevenrm[\string##1]}}}
\def\writeleftlabels{\def\wrlabeL##1{\leavevmode\vadjust{\rlap{\smash%
{\line{{\escapechar=` \hfill\rlap{\sevenrm\hskip.03in\string##1}}}}}}}%
\def\eqlabeL##1{{\escapechar-1%
\rlap{\sixrm\hskip.05in\string##1}%
\llap{\sevenrm\string##1\hskip.03in\hbox to \hsize{}}}}%
\def\reflabeL##1{\noexpand\rlap{\noexpand\sevenrm[\string##1]}}}
\nolabels

\newdimen\fullhsize
\newdimen\hstitle
\hstitle=\hsize % default
\newdimen\hsbody
\hsbody=\hsize % default
\newdimen\hbodyoffset
\hbodyoffset=\hoffset % default
\newbox\leftpage
\def\abstract#1{#1}
\def\rotated{\special{ps: landscape}
\magnification=1000  % This line must come before we change vsize,
                     % since \magnification sets it to a fixed value.
\baselineskip=14pt
\global\hstitle=9truein\global\hsbody=4.75truein
\global\vsize=7truein\global\voffset=-.31truein
\global\hoffset=-0.54in\global\hbodyoffset=-.54truein
\global\fullhsize=10truein
\def\DefineTeXgraphics{%
\special{ps::[global]
/TeXgraphics {currentpoint translate 0.7 0.7 scale
              -80 0.72 mul -1000 0.72 mul translate} def}}
 % 0.7 is slightly less than the ratio of horizontal sizes: 4.75 to 6.5
\let\lr=L
\def\ifsmall{\iftrue}
\def\titlepagefont{\twelvepoint}
\trueseventeenpoint
\def\almostshipout##1{\if L\lr \count1=1
      \global\setbox\leftpage=##1 \global\let\lr=R
   \else \count1=2
      \shipout\vbox{\hbox to\fullhsize{\box\leftpage\hfil##1}}
      \global\let\lr=L\fi}

\output={\ifnum\count0=1 %%% This is the HUTP version
 \shipout\vbox{\hbox to \fullhsize{\hfill\pagebody\hfill}}\advancepageno
 \else
 \almostshipout{\leftline{\vbox{\pagebody\makefootline}}}\advancepageno
 \fi}

\def\abstract##1{{\leftskip=1.5in\rightskip=1.5in ##1\par}} }

% Messages on lines by themselves
\def\linemessage#1{\immediate\write16{#1}}

% tagged sec numbers
\global\newcount\secno \global\secno=0
\global\newcount\appno \global\appno=0
\global\newcount\meqno \global\meqno=1
\global\newcount\subsecno \global\subsecno=0
% and figure numbers
\global\newcount\figno \global\figno=0

\newif\ifAnyCounterChanged
% If we are comparing numbers, there's no special problem.
% But if we are comparing roman numerals, we must be careful, because
% stuff read in from the aux file would be made up of ordinary
% characters (category code = 11), whereas \romannumeral generates
% characters with category code = 12..., so the stuff from the
% current run won't appear equal to the previous definition, as far
% as \warnIfChanged is concerned.
% To get around this, we have a macro \makeNormal, which converts
% letters `ivxlcdmIVXLCDM' to normal letters, no matter what their category
% code.  The macro has the convoluted form it does, with aftergroup's & all,
% to avoid blowing up TeX...
% The macro is used below in makeNormalizedRomappno, by which means we
% define the appendix counters to be strings containing vanilla versions
% of the letters... Sigh
\let\terminator=\relax
% The string to be normalized must not contain { and } tokens...
\def\normalize#1{\ifx#1\terminator\let\next=\relax\else%
\if#1i\aftergroup i\else\if#1v\aftergroup v\else\if#1x\aftergroup x%
\else\if#1l\aftergroup l\else\if#1c\aftergroup c\else%
\if#1m\aftergroup m\else%
\if#1I\aftergroup I\else\if#1V\aftergroup V\else\if#1X\aftergroup X%
\else\if#1L\aftergroup L\else\if#1C\aftergroup C\else%
\if#1M\aftergroup M\else\aftergroup#1\fi\fi\fi\fi\fi\fi\fi\fi\fi\fi\fi\fi%
\let\next=\normalize\fi%
\next}
% makes #1 a normalized version of #2...
\def\makeNormal#1#2{\def\doNormalDef{\edef#1}\begingroup%
\aftergroup\doNormalDef\aftergroup{\normalize#2\terminator\aftergroup}%
\endgroup}
% makes a normalized version of its argument:

\def\warnIfChanged#1#2{%
\ifundef#1% skip it
\else\begingroup%
\edef\oldDefinitionOfCounter{#1}\edef\newDefinitionOfCounter{#2}%
%\message{old: \oldDefinitionOfCounter}%
%\message{new: \newDefinitionOfCounter}%
\ifx\oldDefinitionOfCounter\newDefinitionOfCounter%
\else%
\linemessage{Warning: definition of \noexpand#1 has changed.}%
\global\AnyCounterChangedtrue\fi\endgroup\fi}

\def\Section#1{\global\advance\secno by1\relax\global\meqno=1%
\global\subsecno=0%
\bigbreak\bigskip% (combination \goodbreak\bigskip\bigskip)
\centerline{\twelvepoint \bf %
\the\secno. #1}%
\par\nobreak\medskip\nobreak}
\def\tagsection#1{%
\warnIfChanged#1{\the\secno}%
\xdef#1{\the\secno}%
\ifWritingAuxFile\immediate\write\auxfile{\noexpand\xdef\noexpand#1{#1}}\fi%
}
\def\section{\Section}
\def\Subsection#1{\global\advance\subsecno by1\relax\medskip %
\leftline{\bf\the\secno.\the\subsecno\ #1}%
\par\nobreak\smallskip\nobreak}
\def\tagsubsection#1{%
\warnIfChanged#1{\the\secno.\the\subsecno}%
\xdef#1{\the\secno.\the\subsecno}%
\ifWritingAuxFile\immediate\write\auxfile{\noexpand\xdef\noexpand#1{#1}}\fi%
}

\def\subsection{\Subsection}

\def\romappno{\uppercase\expandafter{\romannumeral\appno}}
\def\makeNormalizedRomappno{%
\expandafter\makeNormal\expandafter\normalizedromappno%
\expandafter{\romannumeral\appno}%
\edef\normalizedromappno{\uppercase{\normalizedromappno}}}
\def\Appendix#1{\global\advance\appno by1\relax\global\meqno=1\global\secno=0%
\global\subsecno=0%
\bigbreak\bigskip% (combination \goodbreak\bigskip\bigskip)
\centerline{\twelvepoint \bf Appendix %
\romappno. #1}%
\par\nobreak\medskip\nobreak}
\def\tagappendix#1{\makeNormalizedRomappno%
\warnIfChanged#1{\normalizedromappno}%
\xdef#1{\normalizedromappno}%
\ifWritingAuxFile\immediate\write\auxfile{\noexpand\xdef\noexpand#1{#1}}\fi%
}
\def\appendix{\Appendix}
\def\Subappendix#1{\global\advance\subsecno by1\relax\medskip %
\leftline{\bf\romappno.\the\subsecno\ #1}%
\par\nobreak\smallskip\nobreak}
\def\tagsubappendix#1{\makeNormalizedRomappno%
\warnIfChanged#1{\normalizedromappno.\the\subsecno}%
\xdef#1{\normalizedromappno.\the\subsecno}%
\ifWritingAuxFile\immediate\write\auxfile{\noexpand\xdef\noexpand#1{#1}}\fi%
}

\def\eqn#1{\makeNormalizedRomappno%
\ifnum\secno>0%
  \warnIfChanged#1{\the\secno.\the\meqno}%
  \eqno(\the\secno.\the\meqno)\xdef#1{\the\secno.\the\meqno}%
     \global\advance\meqno by1
\else\ifnum\appno>0%
  \warnIfChanged#1{\normalizedromappno.\the\meqno}%
  \eqno({\rm\romappno}.\the\meqno)%
      \xdef#1{\normalizedromappno.\the\meqno}%
     \global\advance\meqno by1
\else%
  \warnIfChanged#1{\the\meqno}%
  \eqno(\the\meqno)\xdef#1{\the\meqno}%
     \global\advance\meqno by1
\fi\fi%
\eqlabeL#1%
\ifWritingAuxFile\immediate\write\auxfile{\noexpand\xdef\noexpand#1{#1}}\fi%
}
\def\defeqn#1{\makeNormalizedRomappno%
\ifnum\secno>0%
  \warnIfChanged#1{\the\secno.\the\meqno}%
  \xdef#1{\the\secno.\the\meqno}%
     \global\advance\meqno by1
\else\ifnum\appno>0%
  \warnIfChanged#1{\normalizedromappno.\the\meqno}%
  \xdef#1{\normalizedromappno.\the\meqno}%
     \global\advance\meqno by1
\else%
  \warnIfChanged#1{\the\meqno}%
  \xdef#1{\the\meqno}%
     \global\advance\meqno by1
\fi\fi%
\eqlabeL#1%
\ifWritingAuxFile\immediate\write\auxfile{\noexpand\xdef\noexpand#1{#1}}\fi%
}
\def\anoneqn{\makeNormalizedRomappno%
\ifnum\secno>0
  \eqno(\the\secno.\the\meqno)%
     \global\advance\meqno by1
\else\ifnum\appno>0
  \eqno({\rm\normalizedromappno}.\the\meqno)%
     \global\advance\meqno by1
\else
  \eqno(\the\meqno)%
     \global\advance\meqno by1
\fi\fi%
}
\def\mfig#1#2{\ifx#20%unnumbered figure
\else\global\advance\figno by1%
\relax#1\the\figno%
\warnIfChanged#2{\the\figno}%
\xdef#2{\the\figno}%
\reflabeL#2%
\ifWritingAuxFile\immediate\write\auxfile{\noexpand\xdef\noexpand#2{#2}}\fi\fi%
}

\catcode`@=11 % borrow the private macros of PLAIN (with care)

% \LoadFigure is used to put a figure into the text.  Its first argument
% is the symbolic name for the figure (if it isn't defined, a new number
% will be assigned);  the second argument is a caption;
% the third argument size information in the form
% \epsfxsize=3.0in\epsfysize=3.5in (this argument may be blank and
% may contain any valid preparatory argument used by the epsf package);
% the fourth and last argument is the name of the file which contains the
% figure.
% The macro is basically just a front-end for \epsfbox; its purpose is
% to allow figures to be switched from placement in the running text
% to placement on a separate page at the end of the text.  This choice
% is made using the flag \FiguresInText{true,false}; in the latter case,
% figures are placed at the end, size information is ignored (figures
% will be full-size), and the captions are listed separately on a page
% when the \listfigs command is invoked, followed by the figures, each
% on a separate page.
%  The epsf package must be loaded by the user.
%  To change the size of captions in the text, redefine \captionsize.
\newif\ifFiguresInText\FiguresInTexttrue
\newif\if@FigureFileCreated
\newwrite\capfile
\newwrite\figfile

%default
\newif\ifcaption
\captiontrue
\def\captionsize{\tenrm}
\def\PlaceTextFigure#1#2#3#4{%
\vskip 0.5truein%
#3\hfil\epsfbox{#4}\hfil\break%
\ifcaption\hfil\vbox{\captionsize Figure #1. #2}\hfil\fi%
\vskip10pt}
\def\PlaceEndFigure#1#2{%
\epsfxsize=\hsize\epsfbox{#2}\vfill\centerline{Figure #1.}\eject}

\def\LoadFigure#1#2#3#4{%
%\ifundef#1
\iftrue{
\phantom{\mfig{}#1}}%  Write out definition only if it's new.
\ifx#10% unnumbered figure
\else\warnIfChanged#1{\the\figno}%
\ifWritingAuxFile\immediate\write\auxfile
{\noexpand\xdef\noexpand#1{#1}}\fi\fi\fi%
\ifFiguresInText% Figure is immediate
\PlaceTextFigure{#1}{#2}{#3}{#4}%
\else% Figure is at the end
\if@FigureFileCreated\else%
\immediate\openout\capfile=\jobname.caps%
\immediate\openout\figfile=\jobname.figs%
@FigureFileCreatedtrue\fi%
\immediate\write\capfile{\noexpand\item{Figure \noexpand#1.\ }{#2}\vskip10pt}%
\immediate\write\figfile{\noexpand\PlaceEndFigure\noexpand#1{\noexpand#4}}%
\fi}

\def\listfigs{\ifFiguresInText\else%
\vfill\eject\immediate\closeout\capfile%\parindent=20pt
\immediate\closeout\figfile%
\centerline{{\bf Figures}}\bigskip\frenchspacing%
\catcode`@=11 % borrow the private macros of PLAIN (with care)
\def\captionsize{\tenrm}
\input \jobname.caps\vfill\eject\nonfrenchspacing%
\catcode`\@=\active
\catcode`@=12  % No longer.
\input\jobname.figs\fi}

%\font\titlefont=cmr10 at 16pt
\font\ninerm=cmr9
\font\eightrm=cmr8
\font\sixrm=cmr6

\def\loadtrueseventeenpoint{
 \font\seventeenrm=cmr10 at 17.28truept
 \font\seventeeni=cmmi10 at 17.28truept
 \font\seventeenbf=cmbx10 at 17.28truept
 \font\seventeenit=cmti10 at 17.28truept
 \font\seventeensl=cmsl10 at 17.28truept
 \font\seventeensy=cmsy10 at 17.28truept
}
\def\loadfourteenpoint{
\font\fourteenrm=cmr10 at 14.4pt
\font\fourteeni=cmmi10 at 14.4pt
\font\fourteenit=cmti10 at 14.4pt
\font\fourteensl=cmsl10 at 14.4pt
\font\fourteensy=cmsy10 at 14.4pt
\font\fourteenbf=cmbx10 at 14.4pt
}
\def\loadtruetwelvepoint{
\font\twelverm=cmr10 at 12truept
\font\twelvei=cmmi10 at 12truept
\font\twelveit=cmti10 at 12truept
\font\twelvesl=cmsl10 at 12truept
\font\twelvesy=cmsy10 at 12truept
\font\twelvebf=cmbx10 at 12truept
}

\font\ninei=cmmi9
\font\eighti=cmmi8
\font\sixi=cmmi6
\skewchar\ninei='177 \skewchar\eighti='177 \skewchar\sixi='177

\font\ninesy=cmsy9
\font\eightsy=cmsy8
\font\sixsy=cmsy6
\skewchar\ninesy='60 \skewchar\eightsy='60 \skewchar\sixsy='60

\font\ninebf=cmbx9
\font\eightbf=cmbx8
\font\sixbf=cmbx6

\font\ninett=cmtt9
\font\eighttt=cmtt8

\hyphenchar\tentt=-1 % inhibit hyphenation in typewriter type
\hyphenchar\ninett=-1
\hyphenchar\eighttt=-1

\font\ninesl=cmsl9
\font\eightsl=cmsl8

\font\nineit=cmti9
\font\eightit=cmti8

 % unslanted text italic

\newskip\ttglue
\def\tenpoint{\def\rm{\fam0\tenrm}%
  \textfont0=\tenrm \scriptfont0=\sevenrm \scriptscriptfont0=\fiverm
  \textfont1=\teni \scriptfont1=\seveni \scriptscriptfont1=\fivei
  \textfont2=\tensy \scriptfont2=\sevensy \scriptscriptfont2=\fivesy
  \textfont3=\tenex \scriptfont3=\tenex \scriptscriptfont3=\tenex
  \def\it{\fam\itfam\tenit}\textfont\itfam=\tenit
  \def\sl{\fam\slfam\tensl}\textfont\slfam=\tensl
  \def\bf{\fam\bffam\tenbf}\textfont\bffam=\tenbf \scriptfont\bffam=\sevenbf
  \scriptscriptfont\bffam=\fivebf
  \normalbaselineskip=12pt
  \let\sc=\eightrm
  \let\big=\tenbig
  \setbox\strutbox=\hbox{\vrule height8.5pt depth3.5pt width\z@}%
  \normalbaselines\rm}

\def\twelvepoint{\def\rm{\fam0\twelverm}%
  \textfont0=\twelverm \scriptfont0=\ninerm \scriptscriptfont0=\sevenrm
  \textfont1=\twelvei \scriptfont1=\ninei \scriptscriptfont1=\seveni
  \textfont2=\twelvesy \scriptfont2=\ninesy \scriptscriptfont2=\sevensy
  \textfont3=\tenex \scriptfont3=\tenex \scriptscriptfont3=\tenex
  \def\it{\fam\itfam\twelveit}\textfont\itfam=\twelveit
  \def\sl{\fam\slfam\twelvesl}\textfont\slfam=\twelvesl
  \def\bf{\fam\bffam\twelvebf}\textfont\bffam=\twelvebf%
  \scriptfont\bffam=\ninebf
  \scriptscriptfont\bffam=\sevenbf
  \normalbaselineskip=12pt
  \let\sc=\eightrm
  \let\big=\tenbig
  \setbox\strutbox=\hbox{\vrule height8.5pt depth3.5pt width\z@}%
  \normalbaselines\rm}

\def\fourteenpoint{\def\rm{\fam0\fourteenrm}%
  \textfont0=\fourteenrm \scriptfont0=\tenrm \scriptscriptfont0=\sevenrm
  \textfont1=\fourteeni \scriptfont1=\teni \scriptscriptfont1=\seveni
  \textfont2=\fourteensy \scriptfont2=\tensy \scriptscriptfont2=\sevensy
  \textfont3=\tenex \scriptfont3=\tenex \scriptscriptfont3=\tenex
  \def\it{\fam\itfam\fourteenit}\textfont\itfam=\fourteenit
  \def\sl{\fam\slfam\fourteensl}\textfont\slfam=\fourteensl
  \def\bf{\fam\bffam\fourteenbf}\textfont\bffam=\fourteenbf%
  \scriptfont\bffam=\tenbf
  \scriptscriptfont\bffam=\sevenbf
  \normalbaselineskip=17pt
  \let\sc=\elevenrm
  \let\big=\tenbig
  \setbox\strutbox=\hbox{\vrule height8.5pt depth3.5pt width\z@}%
  \normalbaselines\rm}

\def\seventeenpoint{\def\rm{\fam0\seventeenrm}%
  \textfont0=\seventeenrm \scriptfont0=\fourteenrm \scriptscriptfont0=\tenrm
  \textfont1=\seventeeni \scriptfont1=\fourteeni \scriptscriptfont1=\teni
  \textfont2=\seventeensy \scriptfont2=\fourteensy \scriptscriptfont2=\tensy
  \textfont3=\tenex \scriptfont3=\tenex \scriptscriptfont3=\tenex
  \def\it{\fam\itfam\seventeenit}\textfont\itfam=\seventeenit
  \def\sl{\fam\slfam\seventeensl}\textfont\slfam=\seventeensl
  \def\bf{\fam\bffam\seventeenbf}\textfont\bffam=\seventeenbf%
  \scriptfont\bffam=\fourteenbf
  \scriptscriptfont\bffam=\twelvebf
  \normalbaselineskip=21pt
  \let\sc=\fourteenrm
  \let\big=\tenbig
  \setbox\strutbox=\hbox{\vrule height 12pt depth 6pt width\z@}%
  \normalbaselines\rm}

\def\ninepoint{\def\rm{\fam0\ninerm}%
  \textfont0=\ninerm \scriptfont0=\sixrm \scriptscriptfont0=\fiverm
  \textfont1=\ninei \scriptfont1=\sixi \scriptscriptfont1=\fivei
  \textfont2=\ninesy \scriptfont2=\sixsy \scriptscriptfont2=\fivesy
  \textfont3=\tenex \scriptfont3=\tenex \scriptscriptfont3=\tenex
  \def\it{\fam\itfam\nineit}\textfont\itfam=\nineit
  \def\sl{\fam\slfam\ninesl}\textfont\slfam=\ninesl
  \def\bf{\fam\bffam\ninebf}\textfont\bffam=\ninebf \scriptfont\bffam=\sixbf
  \scriptscriptfont\bffam=\fivebf
  \normalbaselineskip=11pt
  \let\sc=\sevenrm
  \let\big=\ninebig
  \setbox\strutbox=\hbox{\vrule height8pt depth3pt width\z@}%
  \normalbaselines\rm}

\def\eightpoint{\def\rm{\fam0\eightrm}%
  \textfont0=\eightrm \scriptfont0=\sixrm \scriptscriptfont0=\fiverm%
  \textfont1=\eighti \scriptfont1=\sixi \scriptscriptfont1=\fivei%
  \textfont2=\eightsy \scriptfont2=\sixsy \scriptscriptfont2=\fivesy%
  \textfont3=\tenex \scriptfont3=\tenex \scriptscriptfont3=\tenex%
  \def\it{\fam\itfam\eightit}\textfont\itfam=\eightit%
  \def\sl{\fam\slfam\eightsl}\textfont\slfam=\eightsl%
  \def\bf{\fam\bffam\eightbf}\textfont\bffam=\eightbf \scriptfont\bffam=\sixbf%
  \scriptscriptfont\bffam=\fivebf%
  \normalbaselineskip=9pt%
  \let\sc=\sixrm%
  \let\big=\eightbig%
  \setbox\strutbox=\hbox{\vrule height7pt depth2pt width\z@}%
  \normalbaselines\rm}

 % use after $ in ninepoint sections
\def\tenbig#1{{\hbox{$\left#1\vbox to8.5pt{}\right.\n@space$}}}
\def\ninebig#1{{\hbox{$\textfont0=\tenrm\textfont2=\tensy
  \left#1\vbox to7.25pt{}\right.\n@space$}}}
\def\eightbig#1{{\hbox{$\textfont0=\ninerm\textfont2=\ninesy
  \left#1\vbox to6.5pt{}\right.\n@space$}}}

% Page layout
%\newinsert\footins
\def\footnote#1{\edef\@sf{\spacefactor\the\spacefactor}#1\@sf
      \insert\footins\bgroup\eightpoint
      \interlinepenalty100 \let\par=\endgraf
        \leftskip=\z@skip \rightskip=\z@skip
        \splittopskip=10pt plus 1pt minus 1pt \floatingpenalty=20000
        \smallskip\item{#1}\bgroup\strut\aftergroup\@foot\let\next}
\skip\footins=12pt plus 2pt minus 4pt % space added when footnote is present
%\count\footins=1000 % footnote magnification factor (1 to 1)
\dimen\footins=30pc % maximum footnotes per page

\newinsert\margin
\dimen\margin=\maxdimen
%\count\margin=0 \skip\margin=0pt % marginal inserts take up no space
\def\titlefont{\seventeenpoint}
\loadtruetwelvepoint % At FNAL...
\loadtrueseventeenpoint

% \use\cs
% puts in the expansion of `\cs' if it's defined, the literal "\cs" otherwise.
\def\eatOne#1{}
\def\ifundef#1{\expandafter\ifx%
\csname\expandafter\eatOne\string#1\endcsname\relax}
\def\notTrue{\iffalse}\def\isTrue{\iftrue}
\def\ifdef#1{{\ifundef#1%
\aftergroup\notTrue\else\aftergroup\isTrue\fi}}
\def\use#1{\ifundef#1\linemessage{Warning: \string#1 is undefined.}%
{\tt \string#1}\else#1\fi}

%     \ref\label{text}
% generates a number, assigns it to \label, generates an entry.
% To list the refs on a separate page,  \listrefs
% \nref does the same without generating any text at the reference
% point
% June 26 1994: \preref postpones the generation of an entry, along with
% the text, until the first use of the reference

% 09/14/95: Added html...
%\def\hyperref#1#2{\special{html:<a href=\quote#1\quote>}
%{#2}\special{html:</a>}}
% 09/25/95: Now using Tanmoy Battacharya's macros...

%
\catcode`"=11
\let\quote="
\catcode`"=12
\chardef\foo="22
\global\newcount\refno \global\refno=1
\newwrite\rfile
\newlinechar=`\^^J
\def\@ref#1#2{\the\refno\n@ref#1{#2}}
% Added 09/14/95{\the\refno\n@ref#1{#2}}
\def\h@ref#1#2#3{\href{#3}{\the\refno}\n@ref#1{#2}}
\def\n@ref#1#2{\xdef#1{\the\refno}%
\ifnum\refno=1\immediate\openout\rfile=\jobname.refs\fi%
\immediate\write\rfile{\noexpand\item{[\noexpand#1]\ }#2.}%
\global\advance\refno by1}
\def\nref{\n@ref} % Hide to allow redefinitions of \ref,\nref to \preref
\def\ref{\@ref}   % without breaking the latter...
\def\hrref{\h@ref}
% To start a long reference...
\def\lref#1#2{\the\refno\xdef#1{\the\refno}%
\ifnum\refno=1\immediate\openout\rfile=\jobname.refs\fi%
\immediate\write\rfile{\noexpand\item{[\noexpand#1]\ }#2\semi}%
\global\advance\refno by1}
% To continue a long reference...
\def\cref#1{\immediate\write\rfile{#1\semi}}
% To end a long reference...

\def\preref#1#2{\gdef#1{\@ref#1{#2}}}

\def\semi{;\hfil\noexpand\break}

\def\listrefs{\vfill\eject\immediate\closeout\rfile%\parindent=20pt
\centerline{{\bf References}}\bigskip\frenchspacing%
\input \jobname.refs\vfill\eject\nonfrenchspacing}

\def\inputAuxIfPresent#1{\immediate\openin1=#1
\ifeof1\message{No file \auxfileName; I'll create one.
}\else\closein1\relax\input\auxfileName\fi%
}
% For references, some journal names

%and archives...

%\def\hepphref#1{\hyperref{http://xxx.lanl.gov/abs/hep-ph/#1}{archive}%
%{hep-ph/#1}{hep-ph/#1}}

%\def\hepphref#1{\href{http://xxx.lanl.gov/abs/hep-ph/#1}{hep-ph/#1}}

% An .aux file --- for forward references...
\newif\ifWritingAuxFile
\newwrite\auxfile
\def\SetUpAuxFile{%
\xdef\auxfileName{\jobname.aux}%
% Read it in if it exists
\inputAuxIfPresent{\auxfileName}%
% Now write a new one.
\WritingAuxFiletrue%
\immediate\openout\auxfile=\auxfileName}

% Some generally useful notation
\def\L{\left(}\def\R{\right)}
\def\LP{\left.}

\def\LA{\left\langle}\def\RA{\right\rangle}

% Warn about changed counters...
\def\bye{\par\vfill\supereject%
\ifAnyCounterChanged\linemessage{
Some counters have changed.  Re-run tex to fix them up.}\fi%
\end}

\catcode`\@=\active
\catcode`@=12  % No longer.
\catcode`\"=\active

\SetUpAuxFile
\loadfourteenpoint
\input epsf
\overfullrule 0pt
\parskip 5mm
\hfuzz 10pt

\def\pbar{\overline{p}}
\def\atan{\mathop{\rm atan}\nolimits}
\def\jetsum#1{\sum_{#1\in {\rm jet}}}
\def\DO{D\O}

% turning pagenumbers off
\nopagenumbers

\noindent

$\null$

%\vskip -1.6 cm

hep-ph/yyymmdd
\hfill \hfil\break
\rightline{DTP/95/74}
\rightline{Saclay/SPhT--T95/122}

\vskip -2.4 cm

\baselineskip 12 pt
\Title{Recombination Methods for Jets in $p\pbar$ Collisions}

\centerline{\ninerm E. W. N. Glover}
\baselineskip=13pt
\centerline{\nineit Physics Department, University of Durham,
Durham DH1~3LE, UK}
\vglue 0.2cm

\centerline{\ninerm and}
\vglue 0.2cm
\centerline{\ninerm David A. Kosower${}^{\dagger}$}
\baselineskip12truept
\centerline{\nineit Service de Physique Th\'eorique de Saclay,
 Centre d'Etudes de Saclay}
\centerline{\nineit F-91191 Gif-sur-Yvette cedex, France}

\vglue 0.7cm
\centerline{\tenrm Abstract}
\vglue 0.3cm
\vskip -1cm
{\rightskip=3pc
\leftskip=3pc
\tenrm\baselineskip=12pt%\parindent=1pc
\noindent

A jet algorithm must specify how to (re-)combine different partons or
towers into a single four-vector.  Various recombination schemes
have been used experimentally to examine the transverse energy
profile of jets in hadron colliders.
Generally, the data is insensitive to which scheme is used.
However, we argue that the recombination scheme previously used
by the \DO\ collaboration
is expected to have large perturbative corrections
and should not be used for the purposes of making a
quantitative comparison with
fixed-order perturbation theory.
 }
\baselineskip 15 pt

\vglue 0.3cm

\vfil\vskip .2 cm
\noindent\hrule width 3.6in\hfil\break
${}^{\dagger}$Laboratory of the {\it Direction des Sciences de la Mati\`ere\/}
of the {\it Commissariat \`a l'Energie Atomique\/} of France.\hfil\break
\eject

% putting pagenumbers back
\footline={\hss\tenrm\folio\hss}

Hadrons are produced copiously in collisions at
$p\pbar$ colliders.  Those particles carrying the bulk of an event's
energy are usually observed in relatively narrow, collimated sprays,
known as {\it jets}.  To make a quantitative comparison between
theory and experiment using this observation,
one must go beyond a qualitative definition, and give a precise
algorithm for defining a jet.  We must define the experimental
jet algorithm in terms of the measured properties
of hadrons in the detector, and simultaneously define a
theoretical version at the  parton level
to be used in perturbative QCD predictions.

Jet algorithms are not unique, and neither of course are the experimental
results.  Jet definitions in experiments,
broadly speaking, are two-step algorithms
yielding a set of jet axes for a given event along with an assignment
of each particle or calorimeter tower
either to a specific jet, or to no jet at all.
The partonic analog
also yields a set of jet axes and an assignment of partons to specific jets
(or again to no jet at all) for each final-state configuration at the
given order in perturbation theory.
For a jet algorithm to be of any use, however,
differential cross sections using it must be reliably predicted in
perturbation theory. It must therefore
satisfy various criteria, for
example infrared-safety, in order to be calculable sensibly
order-by-order in perturbation theory.

Both the commonly-used cone algorithms and the hadronic version of
the Durham or $k_T$ algorithm
[\ref\HadronicKT{S. Catani, Yu.\ L. Dokshitser, M. H. Seymour, and
B.\ R.\ Webber, Nucl.\ Phys.\ B406:187 (1993)}]
contain a notion of recombination,
wherein the four-momenta of two or more particles are combined to yield
a single four-momentum.  Just as in $e^+e^-$ collisions, there
are various ways to do this.
The theoretically most straightforward way is to treat all initial
particles or partons as massless, and simply add the four-momenta
so that,
$$
E_{T}^{\rm jet} = \jetsum{i} E_{i}\bigg/\cosh \eta^{\rm jet},\anoneqn
$$
and,
$$\eqalign{
\theta^{\rm jet} &=
   \atan \sqrt{\L\jetsum{i} E_{xi}\R^2+\L\jetsum{i} E_{yi}\R^2}\bigg/
                     \jetsum{i} E_{zi},\hskip 10mm\cr
\eta^{\rm jet} &= -\ln\L\tan\L\theta^{\rm jet}/2\R\R,\hskip 10mm
\phi^{\rm jet} = \atan\L\jetsum{i} E_{yi}\bigg/\jetsum{i} E_{xi}\R.
}\eqn\Fourangles$$
The commonly-used
Snowmass algorithm~[\ref\Snowmass{S. D. Ellis, J. Huth, N. Wainer,
K. Meier, N. Hadley, D. Soper, and M. Greco, in {\it Research Directions
for the Decade\/}, Proceedings of the Summer Study, Snowmass,
Colorado, 1990, ed.\ E. L. Berger (World Scientific, Singapore, 1992)}]
adds transverse energies, and forms a
transverse-energy-weighted combination of the rapidities and azimuthal
angles so that
$$
E_{T}^{\rm jet} = \jetsum{i} E_{Ti},\eqn\SnowET
$$
and,
$$
\eta^{\rm jet} = {1\over E_T^{\rm jet}} \jetsum{i} E_{Ti}\eta_i,
   \hskip 1cm
\phi^{\rm jet} = {1\over E_T^{\rm jet}} \jetsum{i} E_{Ti}\phi_i.
\anoneqn$$
The \DO\ collaboration has used a seemingly-similar mixture of these
two approaches, adding the transverse energies as in eqn.~(\use\SnowET), but
defining the
direction by addition of momenta with eqn.~(\use\Fourangles).
While seemingly similar, we shall see that for {\it theoretical\/}
reasons, the \DO\ recombination scheme (unlike the other two) leads to
a poor connection between parton- and hadron-level predictions.

The other important aspect of jet recombination schemes is the
assignment of particles to the jet. Here we should
emphasize that whatever the direction finding algorithm used,
one must attempt to apply
the {\it same\/} assignment algorithm both to experimental data and in
theoretical
calculations if one is to have any hope of making a sensible
comparison. Typically in experimental analyses
fixed cones of radius $R$ are drawn about `seed towers' to determine which
particles
lie within the jet.  A new jet axis is then calculated and the cone moved
until a stable jet center is found.   The net result is that all towers
within radius $R$ of the final jet direction are included in the jet
but the maximal separation between two towers in the jet is $2R$.
Using a fixed cone with one of the partons as a seed tower in a
next-to-leading order calculation, where at most two partons can combine,
forces both partons to lie within a fixed maximal separation $R$ of
each other.  On the other hand, a fixed cone about the final jet
axis in a perturbative calculation corresponds to a variable
maximal separation between the two partons depending on their
relative energies, angles, and rapidities.  It thus corresponds
to the use of a {\it weighted\/} cone containing the two partons.
For example, in the Snowmass algorithm,
the maximal separation between the two partons would be
$$
{E_T^{\rm jet}\over \max(E_{T1},E_{T2})} R
\anoneqn$$
or $2R$ for equal-$E_T$ partons.
The distinction between the two types of cones reflects the
absence of seed towers between the two partons; since this lack of
seed towers between the two partons is an
artefact of  perturbation theory, we would argue that the correct
way to compare theory and experiment is to use a weighted cone
and combine all particles within $R$ of the final jet direction.
Therefore, throughout this paper, we will employ
a weighted-cone algorithm, where two partons are clustered if they
lie within $R$ (chosen to be $1.0$) of the jet axis reconstructed
according to the one of the three recombination algorithms under
consideration.  The next-to-leading order program we
use ~[\ref\Jetrad{W. T. Giele, E. W. N. Glover, and D. A. Kosower,
Phys.\ Rev.\ Lett. 73:2019 (1994)}]
was constructed using the techniques described in
refs.~[\ref\GG{W. T. Giele and E. W. N. Glover,
Phys.\ Rev.\ D46:1980 (1992)\semi
W. T. Giele, E. W. N. Glover, and D. A. Kosower,
  Nucl.\ Phys.\ B403:633 (1993)}]
and the one-loop matrix elements of
ref.~[\ref\EllisSexton{R. K. Ellis and J. C. Sexton,
Nucl.\ Phys.\ B269:445 (1986)}].
While our discussion will be in the context of cone algorithms, the
comments apply to recombination schemes in the $k_T$ algorithm as well.

A recent \DO\ paper~[\ref\Dzero{S. Abachi
et al., Phys. Letts. B357:500 (1995).}] studied the transverse-energy profile
of
jets in certain rapidity and $E_T$ slices.
The \DO\ jet finding algorithm is identical to the Snowmass algorithm,
however the final jet axis was then
reconstructed using the \DO\ recombination scheme;  thus not all
hadrons in the jet will be within the nominal jet radius $R$ of the
jet axis.
The integrated transverse-energy density of a jet of radius
$R$ is given by
$$
\Psi(r) = {\int_0^r\,{dE_T\over dr}\over \int_0^R\,{dE_T\over dr}}
= \LA \sum_{{\rm jets}} {E_T(r)\over E_T(r=R)}\RA_{\rm jets}
\eqn\JetProfile$$
where $E_T(r)$ is the transverse energy within a cone of size
$r=\sqrt{(\Delta\eta)^2 + (\Delta\phi)^2}$ of the jet axis,
and where $\langle\rangle_{\rm jets}$ denotes averaging over all jets
in the given $E_T\times \eta$ bin within the
event sample.
This study claimed to find large discrepancies
between the data and next-to-leading order calculations.

In many of the comparisons in ref.~[\use\Dzero], a fixed-size
cone containing the two partons in an NLO calculation was used as the
clustering criterion, whereas the experimental data analysis used
a Snowmass-type algorithm, which, as we have argued above,
should be matched by a weighted-cone
theoretical clustering
criterion.  Indeed, as shown in fig.~5 of ref.~[\use\Dzero],
an NLO calculation using the Snowmass recombination with a weighted cone
agreed reasonably well with the data (at least outside the jet core)
for both central and forward rapidities.  At smaller values of $r$ the
agreement is less good, but this is understandable since one is then
sensitive to large logarithms, $\ln(r)$.

\topinsert
\def\captionsize{\ninepoint\noindent}
\vskip -2cm
\vbox to 4.5truein{
\vtop to 0pt{
\LoadFigure\JetShapeFigure{\baselineskip 13 pt
\narrower (a) The transverse-energy density profile of jets in the
forward region (b) The same quantity for central jets.
}
{\hskip -3.8truein\epsfxsize 3.5 truein}{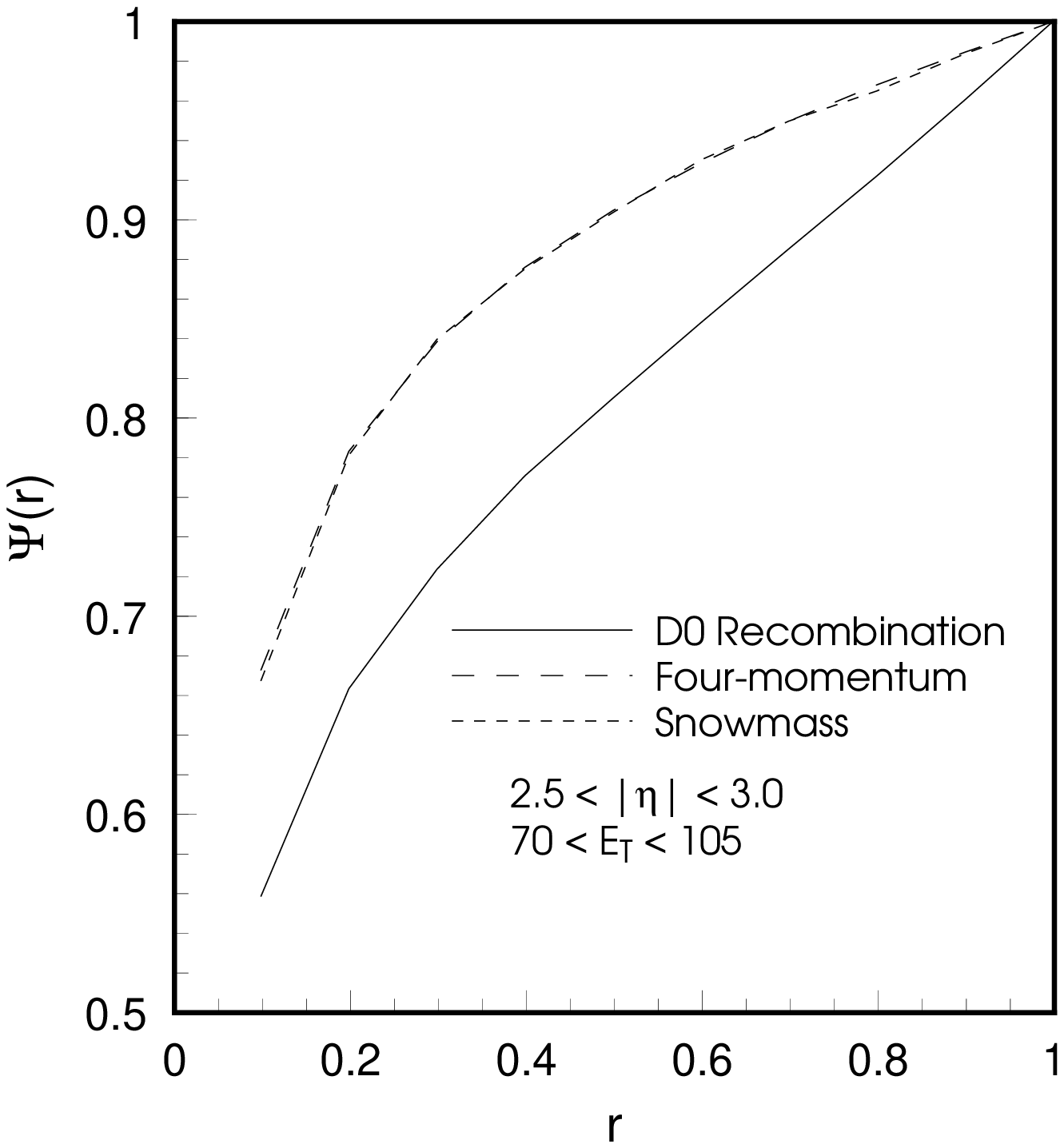}}
\vskip -15pt
\vtop to 0pt{
\captionfalse
\LoadFigure0{}
{\hskip 2.8truein\epsfxsize 3.5 truein}{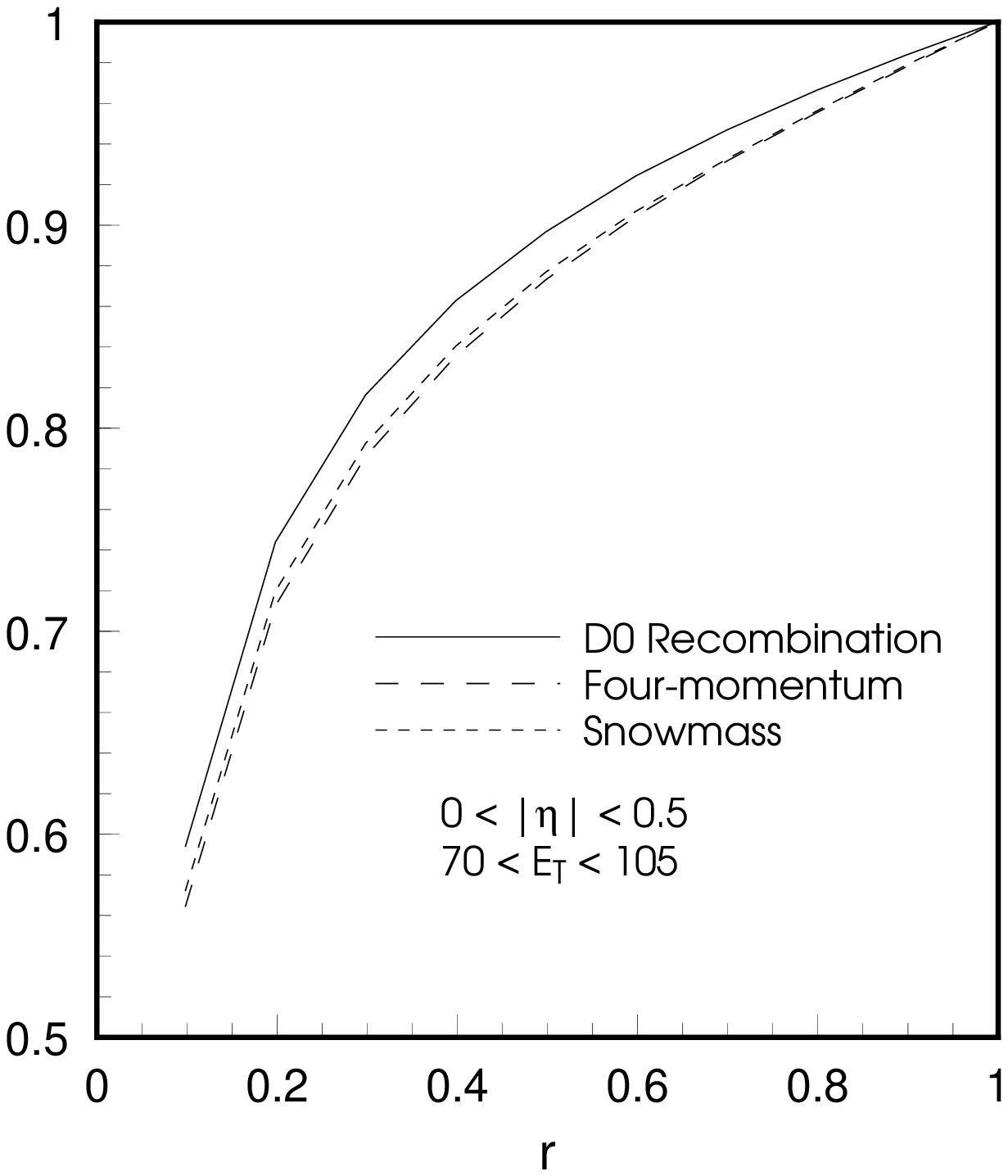}
\captiontrue}}
\endinsert

The same study also showed that the
{\it experimental} measurements of the jet transverse-energy
profile~(\use\JetProfile)
for two different recombination schemes, Snowmass and \DO,
are rather similar. However, it was noted in passing that the
{\it theoretical} predictions
for the jet transverse-energy profile~(\use\JetProfile)
for these two different recombination schemes are
quite different in the forward region (fig.~5 of ref.~[\use\Dzero]).
The same difference is
also shown in figure~\use\JetShapeFigure\ of the present paper.
One may wonder whether this large difference is a result of the
general unreliability of perturbation theory, or whether it is
reflects the difficulties of making a prediction for one of the
two schemes.  As we shall see, it is the latter: quantities
using the \DO\ scheme are poorly predicted in perturbation theory,
and it thus cannot be used for comparing data to predictions from
perturbative QCD.  (Iterative cone algorithms are known to have various
undesirable features from the point of view of fixed-order perturbation
theory; for example, there is typically a prescription for `splitting'
or `merging' partially overlapping jets, configurations that cannot
be modelled at next-to-leading order.  The difficulties with the
\DO\ recombination scheme are additional problems beyond this.)

Any sensible recombination scheme must be infrared-safe: in the limit
where two partons become collinear, or where one parton becomes soft,
it must yield the same jet axis as would be obtained with one fewer
parton.  All three recombination schemes satisfy this constraint.
However, while
it is necessary in order that an observable measured using the
recombination scheme be computable reliably in perturbation theory,
infrared-safety is not sufficient.
\def\xjet{x_{\rm jet}}

To understand the problems with the \DO\ recombination method, it
will be useful to consider the following two quantities, the event's
`jet momentum fraction'
$$
\xjet = \max_{\sigma = \pm} \sum_{{\rm jets}\, j} E_{Tj} e^{\sigma \eta_j}\,,
\eqn\xjetdef$$
\def\defect{\Delta \varepsilon}
and the per-jet fractional energy defect,
$$\eqalign{
\defect &= \LP\L E^{\rm jet} -
             \smash{\sum_{i\in {\rm jet}} }
             \vphantom{\sum_i}
             E_{i}\R\right/\sum_{i \in {\rm jet}}
                            E_{i}\cr
&= \LP\L E_T^{\rm jet}\cosh \eta^{\rm jet}
  - \smash{\sum_{i \in {\rm jet}} }
             \vphantom{\sum_i}
                  E_{T{i}}\cosh \eta_{i}\R \right/
     \sum_{{i}\in {\rm jet}}
                              E_{T{i}}\cosh \eta_{i} \cr
}\eqn\xpartondef$$

Were we to sum over the partons instead of summing over jets in
eqn.~(\use\xjetdef), we would obtain the maximum of the two initial-state
parton momentum fractions $x_{1,2}$, a quantity
that is thus strictly less than one.
The quantity defined of course depends on the jet algorithm used to
cluster partons into jets, and may therefore exceed one.  However,
events with $\xjet > 1$ are guaranteed to have large higher-order
corrections, because they are forbidden at lowest order (where
$\xjet = \max(x_1,x_2)$).

In figure~\use\XJetFigure, we show the distribution in $\xjet$ values.  The
distributions computed using both
the Snowmass and four-momentum recombination schemes
die off as $\xjet\rightarrow 1$,
avoiding at least this source of potentially large higher-order corrections;
the \DO\ recombination scheme, in contrast,
does generate events with $\xjet > 1$.
While the weight of such events may appear to be small in this plot,
in certain regions of phase space, they can lead to dramatic effects.

\topinsert
\def\captionsize{\ninepoint\noindent}
\vskip -2cm
\LoadFigure\XJetFigure{\baselineskip 13 pt
\narrower The single-jet inclusive differential distribution in
$\xjet$, for the forward region.  The lowest-order kinematic boundary
$x=1$ is indicated by a dotted line.}
{\epsfysize  3.5 truein}{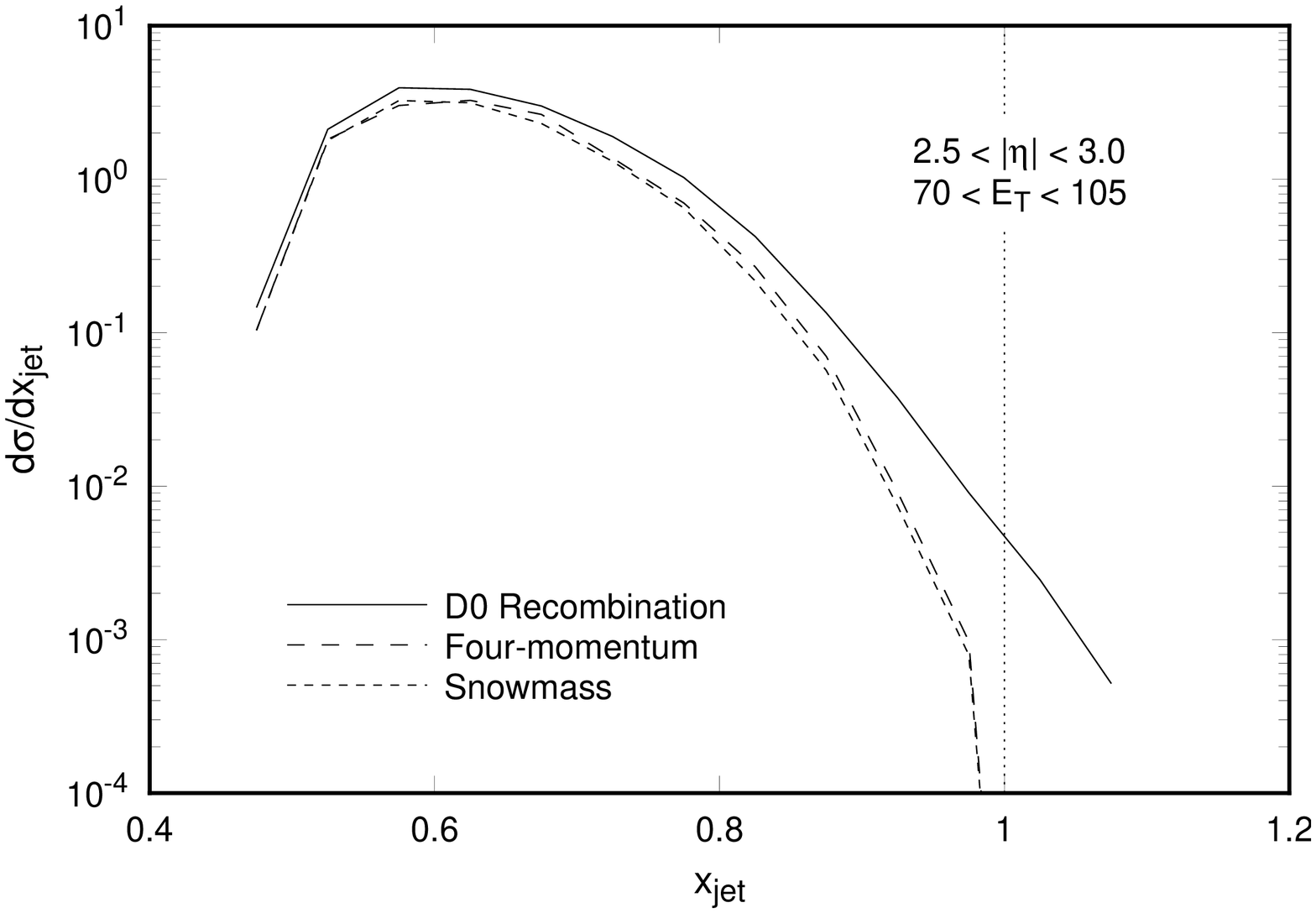}
\endinsert

For events with a jet in the forward region, $\xjet$ is typically
quite close to the total reconstructed energy (scaled by $2/\sqrt{s}$),
and $\xjet>1$ corresponds roughly to events where one jet energy
exceeds one-half the available center-of-mass energy. Since the
jets can be treated as effectively massless, this is
unphysical.

In the \DO\ scheme, the fractional energy defect is given by the
following expression,
$$
\defect =
{(E_{T1}+E_{T2})\sqrt {\L E_{T1} \cosh\eta_1+ E_{T2} \cosh\eta_2\R^2
         +2 E_{T1} E_{T2} \L\cos\Delta\phi-\cosh\Delta\eta\R}
 \over(E_{T1} \cosh\eta_1+ E_{T2} \cosh\eta_2)
   \sqrt{\L E_{T1}+E_{T2}\R^2 + 2 E_{T1} E_{T2} (\cos\Delta\phi-1)}}-1
\eqn\DzeroDefect$$

\def\dphi{\Delta\phi}\def\deta{\Delta\eta}
For small $\dphi$ and $\deta$, this is
$$
\defect \simeq
{E_{T1} E_{T2}\over 2 \L E_{T1}+E_{T2}\R^2}
\L -{\deta^2 \over \cosh^2\eta^{\rm jet}} + \dphi^2 \tanh^2\eta^{\rm jet}\R
+ {\rm higher\ order}
\anoneqn$$

In the forward region, $\cosh\eta^{\rm jet}\gg 1$, while
$\tanh\eta^{\rm jet}\sim 1$, so that this quantity is essentially
positive; furthermore, it increases rapidly as the two partons in
the leading perturbative approximation to the jet move apart in
azimuthal angle.  (One the other hand, in the central region,
$\tanh\eta^{\rm jet} \sim 0$, so that $\defect$ is negative.)
As an example, we can consider a two partons with
equal transverse energies $E_T$, and
$\eta_1=\eta_2 = 2.5$, and $\Delta\phi_{12} = \pi/2$ (with a
cone radius $R=1$, the maximum azimuthal separation for two such
partons within a cluster is $1.778$ radians);
they will be clustered to form a jet of transverse energy $2E_T$
at $\eta = 2.84$.  For such
an event, the fractional energy defect will be $\defect = 0.40$:
the jet's energy will be overestimated by 40\%!
In contrast, the same two partons would reconstruct a jet of transverse
energy $2 E_T$ but $\eta=2.5$ using the Snowmass recombination, or
a jet at $\eta=2.84$ but with transverse energy $\sqrt{2} E_T$ using
four-momentum recombination.
The contribution of such configurations can readily be seen
in figure~\use\DeltaEFigure; with the \DO\ recombination scheme, there is
a substantial tail of events with $\defect > 0$.

\topinsert
\def\captionsize{\ninepoint\noindent}
\vskip -2cm
\LoadFigure\DeltaEFigure{\baselineskip 13 pt
\narrower The single-jet inclusive differential distribution in
$\defect$, for the forward region.  The distribution for the four-momentum
recombination scheme is a delta function at $\defect = 0$, and is not
shown on the plot.}
{\epsfysize 3.5 truein}{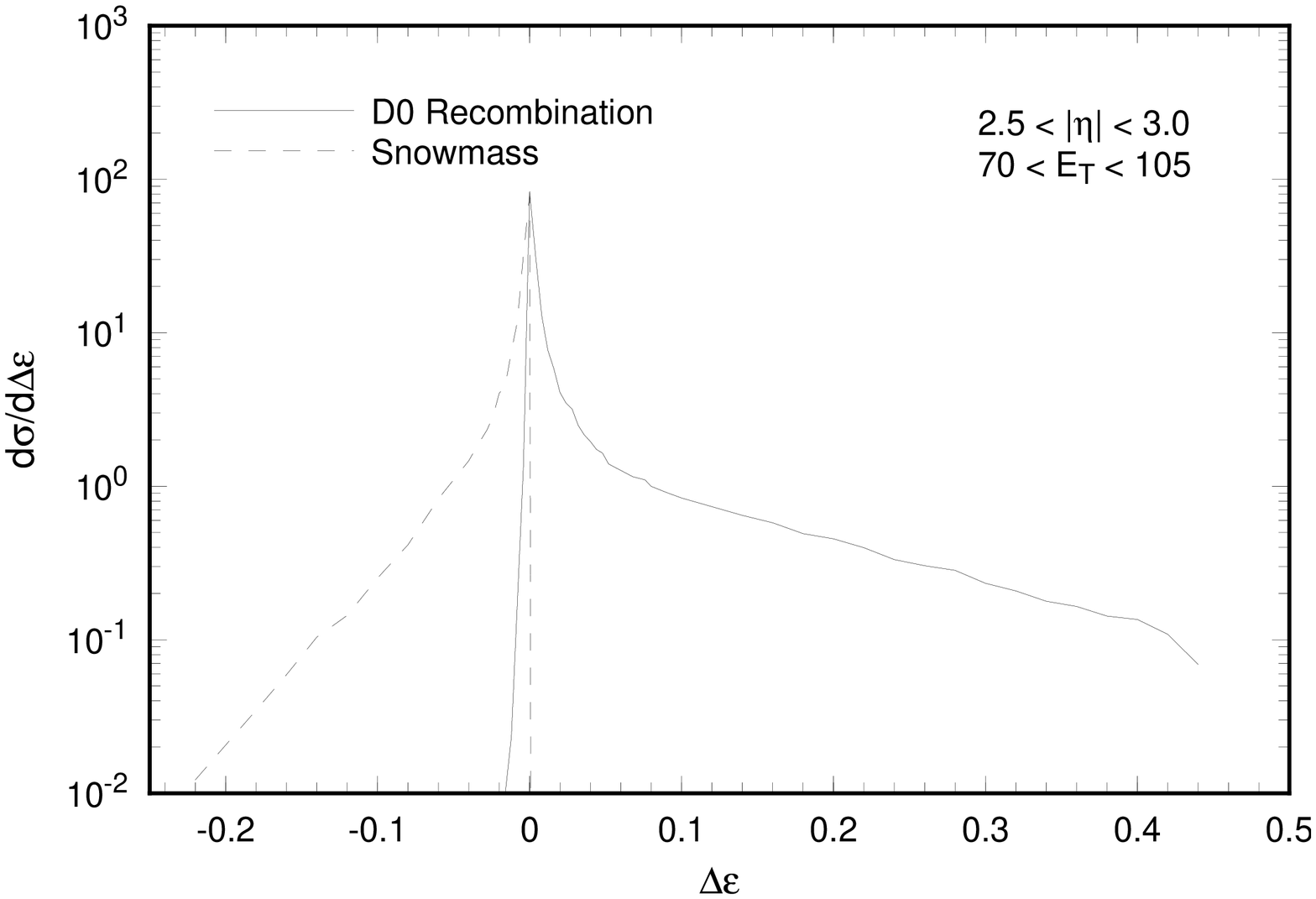}
\endinsert

So long as $\defect \sim 0$ on average, that is on average,
the reconstructed jet energy
is roughly the same as the sum of the parton energies inside it,
one might believe that cross sections and distributions
should not be too different.
However, the parton-level
cross section is falling rapidly as a function of energy,
and thus shifting the assignment of events upwards in energy can
have a dramatic effect.
\iffalse
In particular, events with $\defect > 0$
will show up at higher energy (equivalently, at higher $E_T$ for
fixed $\eta$ or larger $\eta$ for fixed $E_T$).  But compared to events
with the same jet energy but $\defect \simeq 0$, it will have a much
larger weight.  Thus
a given $E_T\times \eta$ bin will
be `contaminated' by contributions with an anomalously large differential
cross-section.\fi
For a given $E_T\times \eta$ bin (and thus for a given reconstructed
jet energy),
events with $\defect>0$ have smaller net partonic energy and therefore
a larger weight, than events with $\defect \sim 0$.
As the analytic results above show, the energy defect in the \DO\
recombination scheme is not only positive in the forward region, but increases
substantially
as the two partons move apart in azimuthal angle.  This is reflected
in the fact that the reconstructed rapidity grows as the two partons
move apart; since the cross section at fixed $E_T$ falls rapidly as
a function of $\eta$ in the forward region, these events will move
to a region of much smaller cross-section, and as a result will carry
a disproportionately large weight.

\topinsert
\def\captionsize{\ninepoint\noindent}
\vskip -2.0cm
\vbox to 4.5truein{
\vtop to 0pt{
\LoadFigure\DeltaEShapeFigure{\baselineskip 13 pt
\narrower (a) The average energy defect as a function of the distance $r$ from
the jet axis, in the forward region. (b) The same quantity in the central
region.}
{\hskip -3.7truein\epsfxsize 3.4 truein}{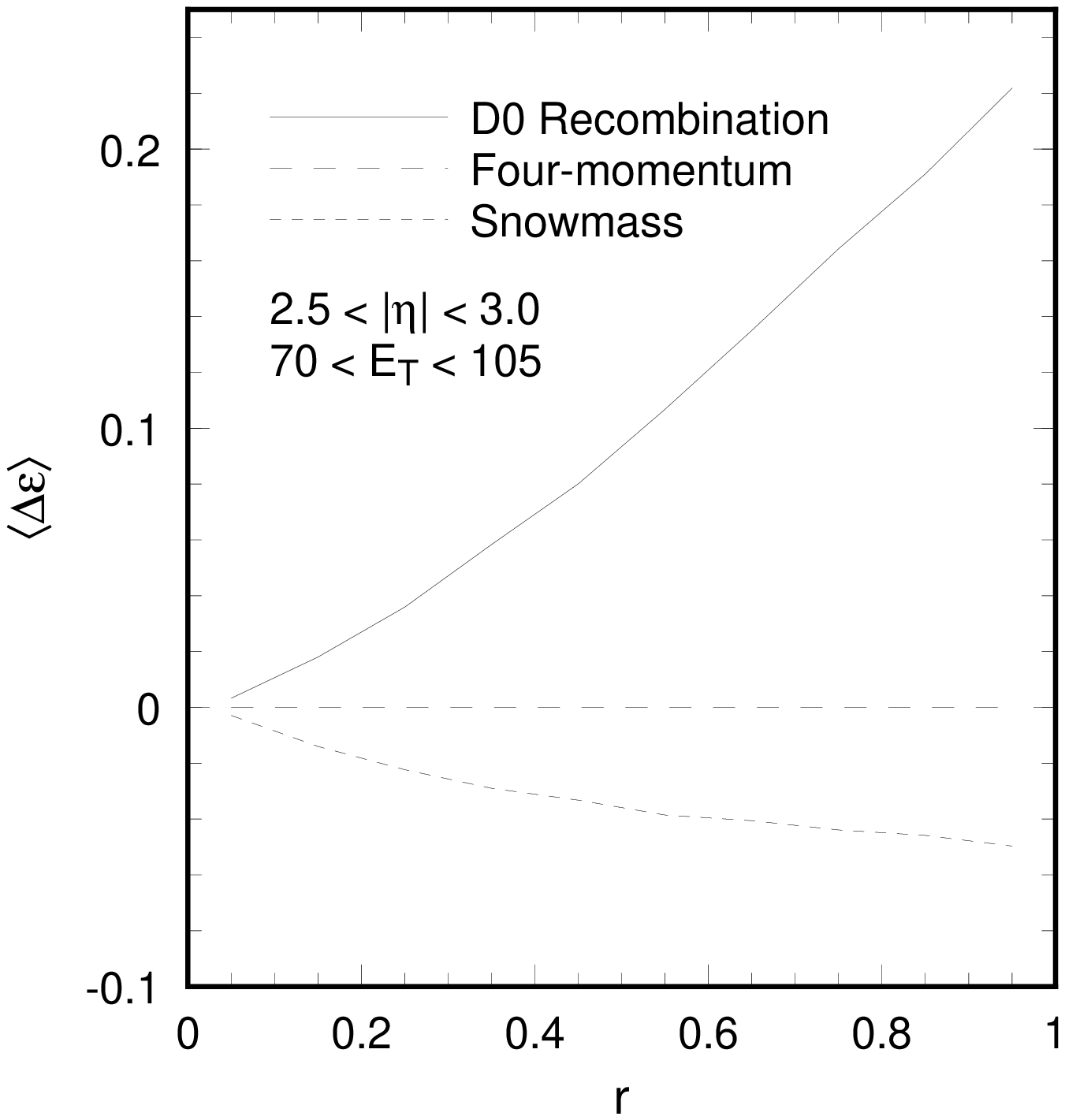}}
\vskip -15pt
\vtop to 0pt{
\captionfalse
\LoadFigure0{}
{\hskip 2.8truein\epsfxsize 3.4 truein}{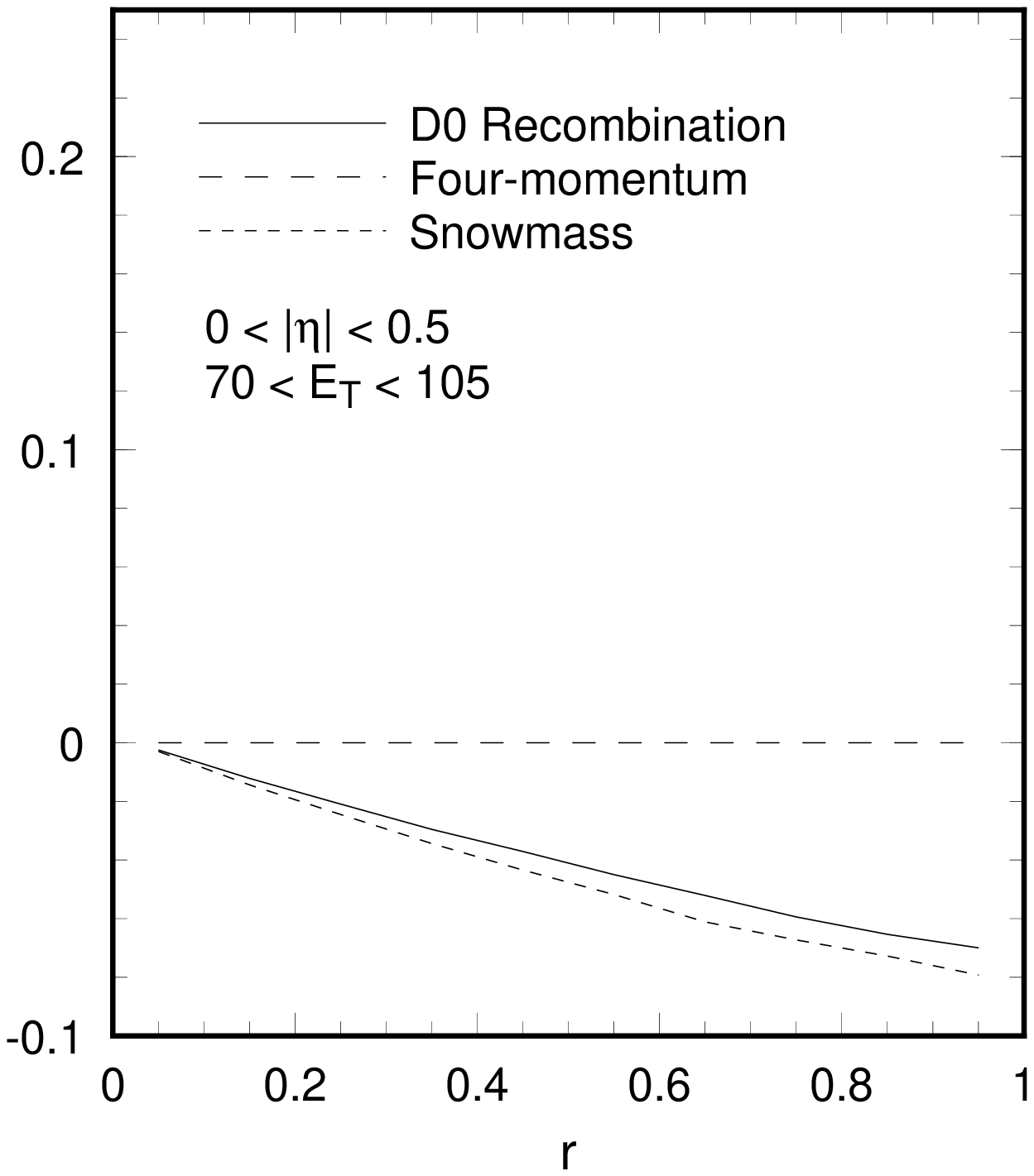}
\captiontrue}}
\endinsert

This increase in the energy defect at large separations can also
be seen in figure~\use\DeltaEShapeFigure(a).
In contrast, the four-momentum recombination scheme
of course has identically zero defect, while the Snowmass scheme
has a negative defect.  In further contrast, as shown in
figure~\use\DeltaEShapeFigure(b), the defects for both
Snowmass and \DO\  recombination
are negative in the central region.
(One might worry about such negative defects;
however, it is physically much more reasonable for an algorithm to
lose energy (for example, through `leakage') than it is for it to find
more energy than was put in.  So long as this negative defect is not
too large or rapidly varying in a distribution, it should not have
much impact on cross sections or distributions.)

The increasing energy defect at increasing separation means that at
larger separation, the differential cross-section within the jet,
or equivalently the transverse energy density within the jet, will
be over-estimated at large $r$, since we will get an
increasing contamination of events with a large weight.  As a
result, at the parton level, jets will be appear much broader.
This is precisely what is seen in figure~\use\JetShapeFigure.

More generally, any distribution across which
the average $\defect$ varies substantially will not be accurately
computed in perturbation theory, since different parts of the
distribution will suffer different degrees of `contamination' from
events of anomalously large weight.  This difficulty is a reflection
of the fact that a next-to-leading perturbative calculation attempts
to model the properties of an average jet
by a weighted sum over a statistical ensemble of two-parton configurations.
(It should be noted, in fact, that such calculations are only next-to-leading
when applied to distributions of the jets in an event; since a leading-order
jet calculation has no internal structure, NLO calculations produce
the leading non-trivial calculation of jet structure.).
In particular, the jet
transverse-energy profile $\Psi(r)$ emerges as the average over a
set of very unsmooth distributions,
$$
\Psi(r) = \LA \sum_{i\in {\rm jet}} {E_{Ti}\over E_{T{\rm jet}}}\delta(r-r_i)
\RA_{\rm jets}
\eqn\Averaging$$
This will
succeed only so long as the different configurations of two-parton
emsembles are treated uniformly by the jet algorithm.  The \DO\
recombination scheme fails this criterion, since two-parton configurations
with large $\Delta r$ are more likely to contain contributions with
$\defect$ substantially larger than zero.

As one goes to higher orders in perturbation theory, one will find jets
with an increasing number of partons.  Although there will be
large jet-to-jet fluctuations about the average,
the transverse-energy profile of
each {\it individual\/} jet will still become less lumpy than at the leading
non-trivial order, and in particular
will contain contributions at many different values of $r$.  This will
tend to smear out the $\defect$ distribution shown in
figure~\use\DeltaEShapeFigure.  On average, that is,
$$
\LA\defect\RA(r) = \LA \defect_{\rm jet}
\sum_{i\in {\rm jet}} {E_{Ti}\over E_{T{\rm jet}}}\delta(r-r_i)
\RA_{\rm jets}
\anoneqn$$
(shown in figure~\use\DeltaEShapeFigure), will increasingly
receive contributions across the whole range of $r$ from each event
as the number of partons in the jet increases; and this will lessen
the variations in $\defect$ as one moves from smaller to larger $r$.
Equivalently, one expects the jet-to-jet fluctuations in $\defect$ to
become smaller, and as this happens, the shape will be less distorted by
the effects considered in this paper; but this of course means that
it may be substantially different from the lowest non-trivial order
prediction given here.  One could quantify these differences by
comparing the sort of NLO calculation considered here, with a calculation
using a parton shower Monte Carlo such as
H{\eightrm ERWIG\/}~[\ref\Herwig{%
G. Marchesini, B. R. Webber. Nucl.\ Phys.\ B310:461 (1988)\semi
G. Marchesini and B.R. Webber, Cavendish preprint HEP--88/7, Oct 1988\semi
G. Marchesini, B. R. Webber, G. Abbiendi, I. G. Knowles, M. H. Seymour,
L. Stanco. Comput.\ Phys.\ Commun.\ 67:465 (1992)}].  We expect
that the difference between an NLO calculation and H{\eightrm ERWIG\/}
 would be small for either the four-momentum
or the Snowmass recombination schemes, but large for the \DO\
recombination scheme.
(Most configurations
produced in parton-shower calculations, or in experimental data, will consist
of a single jet core centered on the eventual jet axis, surrounding by
softer radiation as one moves outward.  Such calculations will also
produce configurations with two widely-separated
jet cores inside the fixed cone.  The
fate of such configurations depends on the prescriptions for `splitting'
and `merging' jets within the jet algorithm.  If they are eventually
classified as single jets, they will distort jet shapes measured using
\DO\ recombination in a manner similar to that found for two-parton
configurations in this paper.  However, since such configurations involve
an additional wide-angle emission, they will be suppressed by a factor
of ${\cal O}(\alpha_s)$, and the distortion if they are retained will
be correspondingly much smaller.
This has been studied by Abbott~[\ref\Brad{%
B. Abbott, Ph.D Thesis, Purdue University (1994) unpublished}]
who showed that the
H{\eightrm ERWIG\/} predictions for the transverse energy
profile using the different schemes are almost identical.)

To summarize,  although the data [\use\Dzero] seem relatively
insensitive to the choice of recombination scheme,
the \DO\ recombination scheme is not as perturbatively stable as
either the four-momentum or Snowmass recombination schemes
and should be discarded for the purposes of making a
quantitative comparison with
fixed order perturbation theory.

We wish to thank the Theoretical Division of CERN, where we studied
this issue,  for its hospitality.  We would also like to thank
Bryan Webber for illuminating discussions about jet algorithms and
higher-order corrections. Finally, we thank Brad Abbott, Jerry Blazey,
Bob McCarthy, Kathy Streets and Harry Weerts for many useful suggestions
and discussions about the \DO\ data.

\listrefs

\bye